%% file: three.tex
\documentclass[tighten,twocolumn]{aastex631}
\usepackage[]{graphicx}
\usepackage{amsmath}
\usepackage[normalem]{ulem}
\newcommand{\hr}{\ensuremath{^\mathrm{h}}}
\newcommand{\mi}{\ensuremath{^\mathrm{m}}}
\newcommand{\msun}{\ensuremath{M_{\odot}}}
\newcommand{\fermi}{\textit{Fermi}}

\shortauthors{Kerr et al.}
\shorttitle{Four $\gamma$-ray Milliseconds Pulsars}

\begin{document}
%\linenumbers

\title{Discovery and Timing of Four $\gamma$-ray Millisecond Pulsars}

\author[0000-0002-0893-4073]{M.~Kerr}
\affiliation{Space Science Division, Naval Research Laboratory, Washington, DC 20375--5352, USA}
\correspondingauthor{M.~Kerr}
\email{matthew.kerr@gmail.com}
\author[0000-0002-7122-4963]{S.~Johnston}
\affiliation{Australia Telescope National Facility, CSIRO, Space and Astronomy, PO Box 76, Epping, NSW 1710, Australia}
\author[0000-0003-4355-3572]{C.~J.~Clark}
\affiliation{
Max Planck Institute for Gravitational Physics (Albert Einstein Institute), D-30167 Hannover, Germany}
\affiliation{Leibniz Universität Hannover, D-30167 Hannover, Germany}
\author[0000-0002-1873-3718]{F.~Camilo}
\affiliation{South African Radio Astronomy Observatory (SARAO), 2 Fir Street,
Observatory, Cape Town 7925, South Africa}
\author[0000-0001-7828-7708]{E.~C.~Ferrara}
\affiliation{Department of Astronomy, University of Maryland, College Park, MD
20742, USA}
\affiliation{Center for Research and Exploration in Space Science and
Technology, NASA/GSFC, Greenbelt, MD 20771, USA}
\affiliation{NASA Goddard Space Flight Center, Greenbelt, MD 20771, USA}
\author[0000-0002-4013-5650]{M.~T.~Wolff}
\affiliation{Space Science Division, Naval Research Laboratory, Washington, DC 20375--5352, USA}
\author[0000-0001-5799-9714]{S.~M.~Ransom}
\affiliation{National Radio Astronomy Observatory, 520 Edgemont Road,
Charlottesville, VA 22903-2475, USA}
\author[0000-0002-9618-2499]{S.~Dai}
\affiliation{Australia Telescope National Facility, CSIRO, Space and Astronomy, PO Box 76, Epping, NSW 1710, Australia}
\author[0000-0002-5297-5278]{P.~S.~Ray}
\affiliation{Space Science Division, Naval Research Laboratory, Washington, DC 20375--5352, USA}
\author[0000-0001-9493-0935]{J.~E.~Reynolds}
\affiliation{Australia Telescope National Facility, CSIRO, Space and Astronomy, PO Box 76, Epping, NSW 1710, Australia}
\author[0000-0003-4337-3897]{J.~M.~Sarkissian}
\affiliation{Australia Telescope National Facility, CSIRO, Space and Astronomy, PO Box 76, Epping, NSW 1710, Australia}
\author[0000-0001-8715-9628]{E.~D.~Barr}
\affiliation{Max-Planck-Institut für Radioastronomie, Auf dem Hügel 69, D-53121 Bonn, Germany}
\author[0000-0002-4175-2271]{M.~K.~Kramer}
\affiliation{Max-Planck-Institut für Radioastronomie, Auf dem Hügel 69, D-53121 Bonn, Germany}
\affiliation{Jodrell Bank Centre for Astrophysics, Department of Physics and Astronomy, University of Manchester, Manchester M13 9PL, UK}
\author[0000-0001-9242-7041]{B.~W.~Stappers}
\affiliation{Jodrell Bank Centre for Astrophysics, Department of Physics and Astronomy, University of Manchester, Manchester M13 9PL, UK}
%\author[]{Aditya}

\begin{abstract}
We discovered four millisecond pulsars (MSPs) in searches of 80 $\gamma$-ray sources conducted from 2015 to 2017 with the \textit{Murriyang} radio telescope of the Parkes
  Observatory.
  We provide an overview of the survey and focus on the results of a
  follow-up pulsar timing campaign.  Using \fermi{} Large Area Telescope data, we have detected $\gamma$-ray
  pulsations from all four pulsars, and by 
  combining radio and $\gamma$-ray data we obtain improved timing
  solutions.  We also provide flux density distributions for the radio
  pulsars and flux-calibrated and phase-aligned radio and $\gamma$-ray
  pulse profiles.  Some of the pulsars may be suitable for radio pulsar timing array experiments. PSR~J0646$-$5455, PSR~J1803$-$4719, and PSR~J2045$-$6837
  are in typical, nearly circular white dwarf binaries with residual
  eccentricities proportional to their binary periods.  PSR~J1833$-$3840 is
  a black widow pulsar with the longest known period, $P_b=0.9$\,d, and a very soft radio spectrum.
  PSR~J0646$-$5455 has a strong, Vela-like $\gamma$-ray pulse profile and is
  suitable for inclusion in the $\gamma$-ray Pulsar Timing Array (GPTA).
  Despite this, it is possibly one of the lowest-efficiency $\gamma$-ray
  MSPs known.  Indeed, all four new $\gamma$-ray MSPs have lower-than-average
  efficiency, a potential indication of bias in earlier searches.  Finally, we retrospectively evaluate the efficiency of this survey: while
  only four new MSPs were directly discovered, subsequent campaigns have
  found pulsars in a further 19 of our targets, an excellent 30\% efficiency.
%\vspace{1cm}
\end{abstract}

%\keywords{}

\section{Introduction}
\label{sec:intro}

Since the launch of the \textit{Fermi} Large Area Telescope \citep[LAT, ][]{Atwood09} in 2008, $\gamma$-ray pulsations have been detected from more than 300 pulsars \citep{Smith23}.  Nearly half of these are
millisecond pulsars (MSPs).  Unlike the most populous
$\gamma$-ray source class, blazars, the emission from pulsars does not
vary on long timescales, and it manifests a nearly universal
spectral shape peaking around 1\,GeV.  It is thus possible to
identify pulsar-like LAT sources and target them for deep and
repeated radio pulsation searches \citep{Ray12}, an approach which has
led to the discovery of well over 100 MSPs.

% Pulsar survey reference?
Targeted observations enable longer integrations compared to wide-area
pulsar surveys, yielding increased sensitivity to faint pulsars.
Repeating the observations yields further advantages.  All pulsars
scintillate in the ionized interstellar medium
\citep[e.g.][]{Narayan92} with concomitant variations in received flux
at the earth.  Observing during a scintillation minimum reduces the
chances of discovering a pulsar, and vice versa.  Furthermore, a
growing subset of MSPs are found in compact binaries, in which ionized
material lifted from the companion surface by the energetic pulsar
wind can eclipse radio emission for substantial fractions of the orbit
\citep[e.g.][]{Polzin20}.  Because the $\gamma$ rays are generally
unaffected unless there is a direct eclipse \citep{Clark23}, targeting
pulsar-like LAT sources has proven to be a particularly effective way
of finding these binaries \citep[e.g.][]{Bangale24}.

Once a pulsar is found in a search of a $\gamma$-ray source, a timing
campaign with a radio telescope is generally needed to establish the
pulsar spin frequency $\nu$, spindown rate $\dot{\nu}$, position, and
Keplerian parameters with sufficient precision to accurately predict
the pulsar spin phase at any given time---i.e. to obtain a timing
solution.  A timing solution enables $\gamma$-ray folding---mapping of LAT photon time stamps
to spin phase---which in turn allows a pulse profile to
emerge from the noise.

Four MSPs were discovered nearly ten years ago in a Parkes
campaign targeting LAT sources, but the follow-up was insufficient to
produce good timing solutions.  These are PSR~J0646$-$5455, previously known as  PSR~J0646$-$54 in the ATNF pulsar catalog \citep[psrcat,][]{Manchester05} and noted as a likely $\gamma$-ray pulsar in the third LAT pulsar catalog \citep[3PC,][]{Smith23}; PSR~J1803$-$4719, not previously in psrcat and listed as PSR~J1802$-$4719 in 3PC; PSR~J1833$-$3840, listed as same in both psrcat and 3PC; and PSR~J2045$-$6837, listed as same in 3PC but as PSR~J2045$-$68 in psrcat. 
 
Because of potential for bolstering the
complement of MSPs available for the $\gamma$-ray Pulsar Timing Array
\citep[GPTA,][]{Ajello22}, we were motivated to obtain new data from the
\textit{Murriyang} radio telescope at the Parkes Observatory.  With it,
archival data, and new techniques for phase connection, we have obtained
good timing solutions and detected $\gamma$-ray pulsations from all four
MSPs.  The detection and pulse profile for J1833$-$3840 were already
reported in 3PC, but we provide a characterization of its radio properties and a much-improved timing solution and $\gamma$-ray pulse profile.

Below, we present an overview of the unpublished details of the
pulsar search campaigns which detected the MSPs (\S\ref{sec:disco});
describe the original and updated radio timing campaigns
(\S\ref{sec:obs_radio}); discuss the detection of $\gamma$-ray
pulsations and improved timing thereby enabled; present
multi-wavelength properties (\S\ref{sec:obs_mwl}); and discuss the
resulting timing solutions, pulse profiles, suitability for
high-precision timing work, and the retrospective survey efficiency (\S\ref{sec:discussion}).

\section{Radio Search Campaign}
\label{sec:disco}

Targeted observations of LAT sources were conducted over several
iterations of the observing program P814 with the 64-m Parkes radio
telescope (now \textit{Murriyang}\footnote{``Skyworld'' to the
Wiradjuri people.}).  \citet{Kerr12} report results from an initial
survey of sources in the 1FGL $\gamma$-ray source catalog
\citep{Abdo10_1FGL}, while \citet{Camilo15} discuss searches
of 2FGL sources and consider detection statistics.
% 10 MSPs in Camilo15
% I guess Elizabeth's sources were based on 3FGL?
Further searches based on a preliminary version of the 4FGL
\citep{Abdollahi20} catalog were conducted from 2015 to 2017.  As with
previous searches, these sources were selected by their lack of association
with other known source classes; their lack of variability; and a spectral
shape similar to those of other pulsars, viz. a power law with a cutoff at
roughly 1 GeV; see \citet{Camilo15} for more details.  All data taken in
2015 and 2016, on 65 unique targets, were reduced soon after observation
and were searched for pulsation as described in \citet{Kerr12} and \citet{Camilo15}.  The final data, from 2017, included 15
new targets that were processed recently, as described below.

All observations were carried out at 20\,cm using the DFB4 backend at a
center frequency of 1369\,MHz with 256\,MHz of bandwidth filtered into
512-channel total-power spectra recorded every 80\,$\mu$s.  From October
2015 through March 2016, we used the multi-beam receiver.  We used the H-OH
receiver for the remainder of 2016, and we resumed use of the multi-beam
receiver in 2017.  Observations were typically one hour in duration, with
shorter observations filling schedule gaps.  The target names (which reflect the position), precise pointing information, observation dates, and integration times are available in Appendix \ref{sec:targets}.  Here, we summarize the observing program and discoveries.

% NB from Fernando's email: October targets, 17 observations of 12
% John S and I did the obs.
% P8Y6 sources, J1832-38 turned up.
% NB two other good looking candidates, 
% J1204-5033
% J1805-3619 (P8Y6J1803)
In 2015, 13 targets were observed, most more than once.
PSR~J1833$-$3840 was discovered in
the first attempt on 2015 October 17, and was later confirmed in a short
observation with the Green Bank Telescope.

% The Feb 2016 observations, from an email from Fernando, featured
% repeated observations of "my" sources, as well as new sources from
% Elizabeth.
In the second half of the semester, February and March of 2016, 14 additional
new targets were observed, and PSR~J1803$-$4719 was discovered in the first observation of the source,
on 2016 February 3.  PSR~J0646$-$5455 was discovered in a 60-minute
observation carried out on 2016 March 28\footnote{PSR~J0646$-$5455
and PSR~J2045$-$6837 may be the first and only pulsars discovered with
the H-OH receiver.}.  The pulsar had not been
detected in two February observations of 30- and 60-minute duration,
likely due to scintillation.

In the first half of the October 2016 semester, we observed 38 new
targets.  Most sources were observed only once, for one hour.
PSR~J2045$-$6837 was discovered on 2016 November 6.

Finally, in February of 2017, we observed 15 new targets.  We reduced
those data using \texttt{PRESTO} \citep{Ransom01} as follows: 
we used \texttt{rfifind} to identify
radio frequency interference (RFI) and blank the affected time and
frequency bins; we divided data into 8 sub-bands and dedispersed them
into time series with trial DMs ranging from 0 to 500\,pc\,cm$^{-3}$; and we
performed an \texttt{accelsearch} with \texttt{zmax}$=200$ and
summing up to 8 harmonics.  We did not find any additional pulsar candidates, though some of these targets are now known to be pulsars (see \S\ref{sec:retro}).

\begin{figure*}
\centering
%\vspace{0.2cm}
  \includegraphics[angle=0,width=0.98\linewidth]{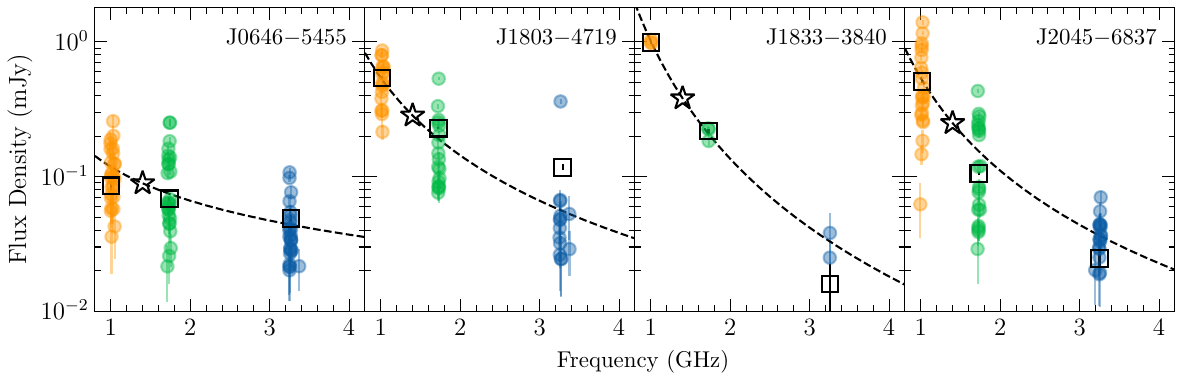}
  \caption{\label{fig:fluxes}Colored points give the measured flux
  density for each UWL epoch, with fold mode measurement preferred
  when available.  The sub-bands are centered on 1.0\,GHz (orange), 1.7\,GHz (green), and 3.3\,GHz (blue).  The dashed line indicates the best-fit power-law
  flux model to the mean flux densities, with the black stars using this model to estimate the flux density at the canonical 1.4\,GHz.  The black squares give the mean flux density for the fold-mode data only---i.e. just for latter epochs----and are shown here for comparison because these observations are used to estimate the timing precision (see Table \ref{tab:timing} and \S\ref{sec:radio_ptas}).}
\end{figure*}

\section{Radio Timing Campaign}
\label{sec:obs_radio}

%\begin{center}
%\begin{table}
%\begin{tabular}{ |l|r|r|}
%PSR & P789/P814 Epochs & P1229 Epochs \\
%\hline
%\hline
%J0646$-$5455 & 13 & ? \\
%J1803$-$4729 & 33 & ? \\
%J1833$-$3840 & ? & ? \\
%J2045$-$6837 & 4 & ? \\
%\hline
%\end{tabular}
%\caption{\label{tab:obs}Lorem ipsum.}
%\end{table}
%\end{center}

In a typical follow-up campaign to obtain a pulsar timing solution,
observations are scheduled with a quasi-logarithmic spacing, e.g. at
separations of 0.5\,d, 1\,d, 2\,d etc., saturating at a monthly
monitoring cadence.  With this bootstrap approach, the measured and
required precisions on parameters like $\nu$ increase at roughly the
same rate.  The goal is to predict the pulse phase at the next
observing epoch to within at most a few neutron star rotations
(wraps).  As the data span lengthens, parameters like the spindown
($\dot{\nu}$) and position become measurable.  Binarity complicates the
picture, especially if the orbital period has a similar timescale to
other parameters.  With about one year of data, it is generally
possible to obtain a good timing solution, historically by inspired
guessing of the number of wraps between observing epochs.

More recent methods aim to systematize and automate this guessing
procedure \citep{Freire18,Phillips20,Taylor24}.  Even with such
automated tools, a timing solution can be difficult to obtain in
practice.  It may not be possible to schedule observations optimally,
and---due to scintillation, eclipses, radio-frequency interference
(RFI), or telescope outages---some epochs may not yield a pulsar
detection.  Obtaining a unique timing solution using brute force
methods rapidly becomes intractable with sparse data.

All four MSPs were observed both with residual P814 time and under the
auspices of P789, a catch-all program for follow-up timing of MSPs
\citep[e.g.][]{Spiewak20}.  %The number of epochs for each pulsar is
%given in Table \ref{tab:obs}.
Observations were carried out at a typical frequency of 1.4\,GHz and a
length of 30--60 minutes.  Due to their later discovery, the coverage
for J0646$-$5455 and J2045$-$6837 was insufficient to yield a timing
solution.  And while more than 30 epochs for J1803$-$4729 and
J1833$-$3840 were obtained, there were too few optimally-spaced epochs
to obtain a timing solution using available methods at the time.

To aid the process of obtaining a unique timing position, we observed one of the new MSPs, PSR~J2045$-$6837, with MeerKAT using the $L$-band ($856$--$1712$~MHz) receiver for 10 minutes on 24 August 2021 and 31 August 2021, using the FBFUSE and APSUSE instruments to form and record coherent tied-array total-intensity beams. In the first observation, we formed 480 beams covering the \textit{Fermi} localization region, and searched these for pulsations around the known DM using the \texttt{Peasoup} pulsar searching software \citep{Barr2020+peasoup}. In the second observation, we used a denser tiling of 27 beams around the position of the brightest detection from the first observation. The pulsar was detected in 23 out of 27 of these beams with a peak signal-to-noise ratio exceeding 80. From these detections on the second epoch, we obtained a interferometric position with approximate $1^{\prime\prime}$ uncertainties using the \texttt{SeeKAT} software \citep{Bezuidenhout2023+SeeKAT}. 

While preparing the proposal for what would become P1229, we attempted
to solve these pulsars again.  We were able to obtain a timing solution for
J1833$-$3840 using new automated tools \citep{Phillips20,Taylor24} and
subsequently detected $\gamma$-ray pulsations \citep{Smith23}.
Consequently, we only observed three MSPs for P1229 in a one-year
campaign beginning in November 2023.

For P1229, we used the Ultra Wide-bandwidth Low \citep[UWL,][]{Hobbs20} receiver to collect coherently dedispersed search-mode data over the 704--4032\,MHz band, sampling 3328-channel total power spectra every 64\,$\mu$s.  Typical observations were
one hour in length, and in some cases we observed the pulsed noise diode
calibration source beforehand.  Once we obtained initial timing solutions (see below), we began to collect simultaneous search- and fold-mode data, the latter having 1024 phase bins and all four Stokes parameters.

\subsection{PRESTO analysis}
We reduced the data in two different ways.  We processed all search
mode data with PRESTO \citep{Ransom01}, using \texttt{rfifind} for RFI
mitigation; \texttt{prepdata} to produce barycentered time series;
and \texttt{accelsearch} to identify candidate pulsed signals near the
known rotation frequencies.  We folded the observations with
detections and extracted pulse times-of-arrival (TOAs) with a simple
gaussian template.  We did this for each of three sub-bands (see \S\ref{sec:psrchive}). We reduced the archival DFB4 timing data in the same
way but did not further divide the 256\,MHz bandwidth.

To obtain timing solutions, we developed a new technique building on
the approaches of \citet{Freire18}, \citet{Phillips20} and \citet{Taylor24} 
but exploiting the correlations between phase wraps and timing model 
parameters to avoid expensive timing-model evaluations, and 
incorporating prior information to reduce the search space, speeding 
up the phase-connecting process by several orders of magnitude (Clark, C. J. 
and van Haasteren, R., in prep.).  We further improved on this by 
adopting lattice reduction techniques, inspired by \citet{Gazith24}.
With this approach, we obtained timing solutions to all three MSPs. 
While the solutions for PSRs J1803$-$4719 and J2045$-$6837 were unambiguous, 
the radio TOAs for J0646$-$5455 yielded several candidate phase-connected timing
solutions with different rotation counts between widely separated TOAs. 
Only one of these yielded $\gamma$-ray pulsations when we used it to fold the
\textit{Fermi}-LAT data. With these  timing solutions, we used \texttt{PRESTO} to produce a final set of TOAs using multi-gaussian models of the pulse profiles.  
These are the radio TOAs used for all timing solutions.

\subsection{psrchive analysis}
\label{sec:psrchive}
To obtain calibrated flux density measurements and detailed pulse profiles,
we used \texttt{psrchive} \citep{Hotan04}.  First, we folded the search
mode data using $\texttt{dspsr}$ \citep{vanStraten11}.  We then processed
the folded profiles and the native fold-mode data using a pipeline \citep{Johnston21}
developed for the Parkes young pulsar timing program P574
\citep{Weltevrede10}.  During this step, polarization and flux calibration are
applied.  For most epochs, we have at least one contemporary observation of
the calibration source, while for a few, we use the nearest-in-time
calibrator.  The UWL gains are stable for weeks, but change substantially
after a system reset.  We checked that the epochs with missing calibrators
appeared to be in a stable phase well represented by the close-in-time
calibration solution.

We further split the calibrated data into three sub-bands covering the
frequency ranges 704--1216\,MHz, 1216--2240\,MHz, and 2496--4032\,MHz,
excluding the two 128\,MHz sub-bands near the 2.4\,GHz WiFi band due
to severe RFI contamination.  Each of the remaining bands is also
affected by RFI, which we manually excise, with roughly 50\%, 30\%,
and 15\% of the frequency-time samples in a typical observation lost.

We built a single analytic model of the pulse profile for each pulsar
by fitting multiple gaussian components to a single high-S/N
observation from the mid-frequency sub-band.  Using the calibrated
profiles, this standard pulse profile model, and the \texttt{psrflux}
tool, we computed the flux density $S_{\nu}$ in each sub-band at each
epoch.  We discarded epochs with a fractional uncertainty $>$0.33 and show the resulting values in Figure \ref{fig:fluxes}.  We fit a power-law model, $S_{\nu}=S_{1400}\,(\nu/1400\,\mathrm{MHz})^{\alpha}$, to these data and report the best-fit parameters in  Table \ref{tab:timing}.

\begin{figure}
    \includegraphics[angle=0,width=1.00\linewidth]{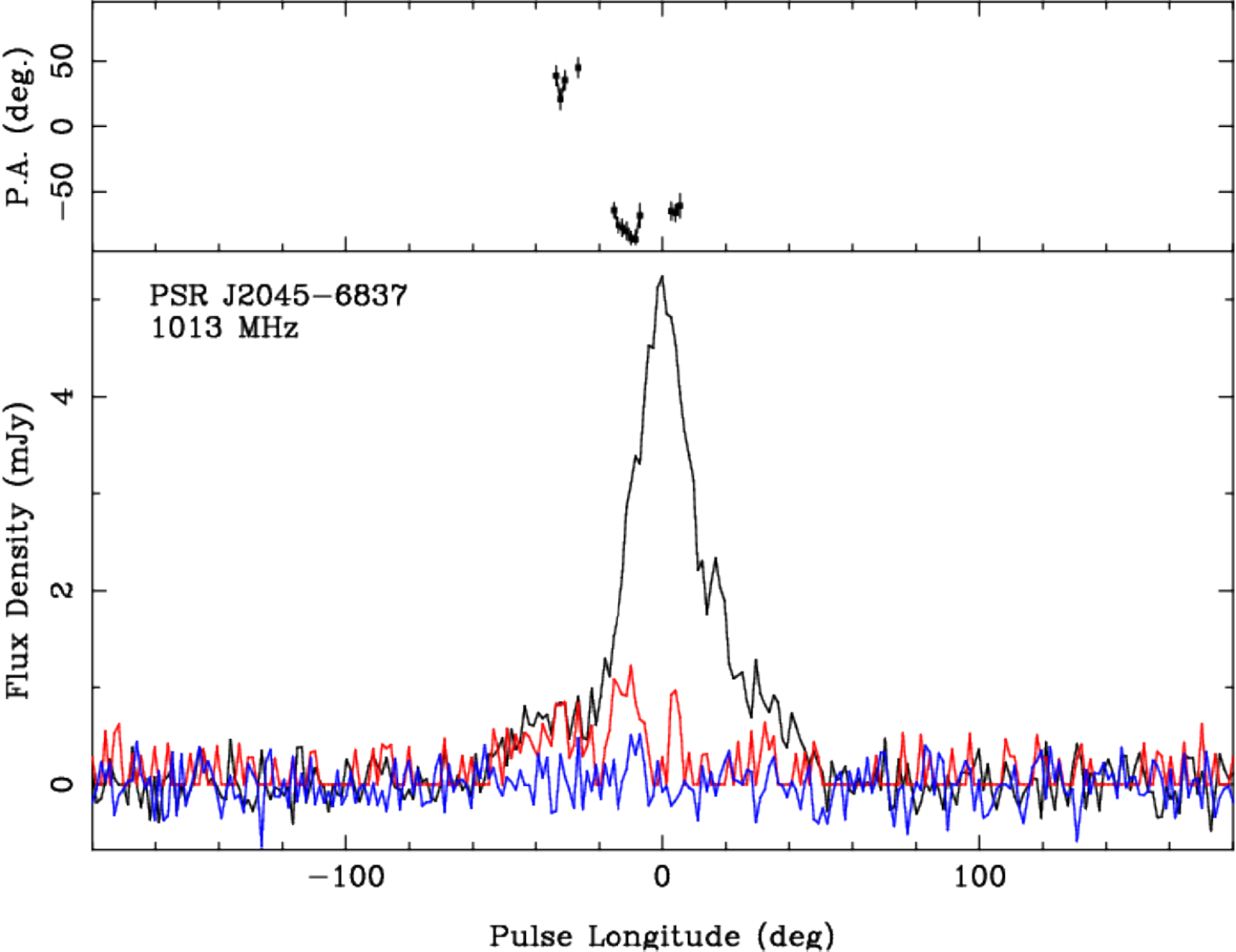}
  \caption{\label{fig:j2045_pol}The polarization properties of
  PSR~J2045$-$6937 in the low-frequency sub-band.  The black, red, and
  blue traces gives Stokes I, $L=U+Q$, and V, respectively.  The
  left shoulder appears to be nearly 100\% linearly polarized,
  allowing measurement of the polarization position angle (P.A.) for some phase bins.}
\end{figure}

We used the timing solutions to co-add the fold-mode data and produce
higher S/N pulse profiles.  With these, we searched for evidence of frequency-dependent polarization rotation in the ISM with the \texttt{rmfit} tool.  Only PSR~J2045$-$6837 had detectable linear polarization, and we measured rotation measure RM = $17 \pm 3$\,rad\,m$^{-2}$.  The de-rotated, frequency-averaged polarization profile appears in Figure \ref{fig:j2045_pol}.

\input{timing_table.tex}

\begin{figure*}
  \includegraphics[angle=0,width=1.00\linewidth]{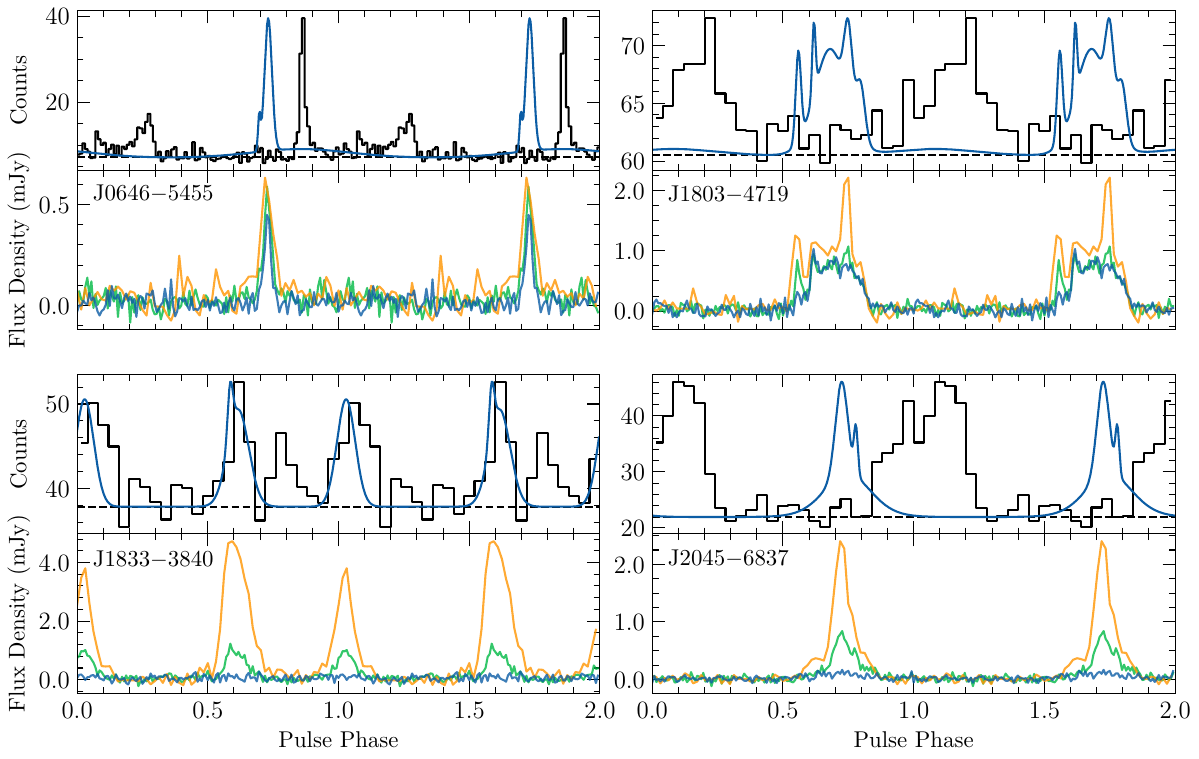}
  \caption{\label{fig:pulse_prof}Phase-aligned $\gamma$-ray and radio
  pulse profiles.  The top panels for each pulsar show histograms of
  the LAT photon weights, viz. the $\gamma$-ray pulse profiles, with
  the dashed black line indicating the estimated background level \citep[see][]
  {Smith23}.  The smooth blue curve shows the analytic radio
  template for reference.  The bottom panels for each pulsar show the
  Stokes I pulse profile for the
  co-added fold-mode data acquired in each of the three sub-bands:
  yellow is 704--1216\,MHz, green is 1216--2240\,MHz, and blue is
  2496--4032\,MHz.}
\end{figure*}

\section{$\gamma$-ray Analysis}
\label{sec:obs_gamma}

We used Pass 8 \citep{Atwood13,Bruel18} data collected from MJD 54682--60562
(16.1\,yr)
with reconstructed positions lying within 3\degr of the pulsar
positions and with reconstructed energies ranging from 100\,MeV to
30\,GeV.  To enhance the precision of the timing analysis
\citep{Bickel08,Kerr11}, we used the 4FGL-DR4 \citep{4FGL_DR4} sky model to compute photon
weights, i.e. the probability that each photon originates from the
pulsar.  We used the radio-data only timing solutions as input and
analyzed the data with a $\gamma$-ray pulsar timing pipeline described
further in \citet{Ajello22}.

For this work, we implemented a new capability in the $\gamma$-ray
timing pipeline: in addition to the unbinned Poisson likelihood, we
use informative gaussian priors provided by the central values and
1-$\sigma$ uncertainties from the radio ephemeris.  The resulting
timing solution thus includes all available constraints, remains
consistent with the radio TOAs, and benefits from the much longer time
span of the $\gamma$-ray data.  This long observing span $t$ is especially effective in determining parameters whose effects accumulate over time, like proper motion ($\propto t$) and $\dot{\nu}$ ($\propto t^2$).

In the case of PSR~J1803$-$4719, the $\gamma$-ray source is faint and the
extensive P814 radio data allow good measurement of position,
Keplerian, and spin-down parameters.  Consequently we report the results of
radio-only timing.  Conversely, for PSR~J0646$-$5455 and PSR~J2045$-$6837,
the posterior is dominated by the $\gamma$-ray data.  For all four
pulsars, the best timing solutions are given in Table \ref{tab:timing} and
provided in the online material\footnote{\url{https://fermi.gsfc.nasa.gov/ssc/data/access/lat/ephems/}}.  We used these timing solutions to fold
the $\gamma$-ray data and, accounting for dispersion and light travel time,
prepared the joint $\gamma$-ray and radio pulse profiles shown in Figure
\ref{fig:pulse_prof}.

The spectral properties of the $\gamma$-ray sources are available in
the 4FGL catalog, and we report in Table \ref{tab:timing} the relevant
spectral parameters from a power-law exponential cutoff model (see 3PC
for an extensive discussion).  PSR~J0646$-$5455 and J2045$-$6837 have relatively typical $\gamma$-ray MSP spectra, with a $d_\mathrm{p}$ (see caption of Table \ref{tab:timing} for a definition) value approaching the limit of $4/3$ for monoenergetic curvature radiation (see 3PC).  On the other hand, the spectrum of PSR~J1803$-$4719 is unusually soft.  Given an estimate of distance from the DM using the model of
\citet{YMW17}, we obtain $\gamma$-ray efficiencies of 1--7\%,
substantially below the typical values of the MSP population reported
in 3PC.  We discuss this further below.

\section{Other Wavelengths}
\label{sec:obs_mwl}

\subsection{X-rays}

The fields of all four MSPs have been observed with the XRT \citep{Burrows05} on the Neil Gehrels Swift observatory with the following livetimes: J0646$-$5455, 0.8\,ks in one ObsID; J1803$-$4719, 3.4\,ks in 8 ObsIDs; J1833$-$3840, 7.2\,ks in 12 ObsIDs; and J2045$-$6837 for 4.7\,ks in 13 ObsIDs.  We examined the resulting images and, within the 18'' half-power diameter of the XRT point-spread function, found respectively 0, 0, 1, and 2 events.  These rates are consistent with the expected background levels.  Detecting a potential X-ray counterpart will thus require longer observation with a more sensitive X-ray telescope.

\subsection{Optical}

The MSP fields are not in the footprints of any of the deep southern sky surveys, and there are no counterparts listed in the third
Gaia data release \citep{Gaia_DR3}.  For white dwarfs, this catalog is
substantially complete only to about 100\,pc \citep{Tremblay24}, much closer than
the DM distances for any of the systems.  It is perhaps more surprising that the high-$\dot{E}$ black widow J1833$-$3840 remains undetected given similar (though closer) systems with optical magnitudes $<$20.  However, we modeled the predicted r-band magnitude using Icarus \citep{Breton12}, assuming an inclination of 60\arcdeg, a pulsar mass of 1.6\msun, 25\% irradiation efficiency, and with the companion star filling its Roche lobe.  The resulting companion temperature peaks at 4300\,K on the day side, and at the DM distance, gives a peak magnitude of r=22.6, well fainter than the Gaia limiting magnitude.

\section{Discussion}

Before synthesizing the results of the survey, we first discuss some
interesting aspects of each of the new pulsars.
\label{sec:discussion}

\subsection{PSR~J0646$-$5455}

This pulsar has by far the highest $\gamma$-ray signal-to-noise in the sample, concordant with a DM-estimated distance of only 0.4\,kpc.  However, with this proximity, the measured efficiency is $<$1\%, one of the lowest
values ever measured for a $\gamma$-ray MSP.  The implied
tangential velocity is well constrained and small: 8.4$\pm$0.4 km/s.  Taken
together, these two anomalies suggest that the distance may be
underestimated, and indeed the NE2001 implied distance is 1.8\,kpc, the largest discrepancy with YMW16 model distances in the sample.  (See \citet{Olszanski25} for a recent discussion of differences between these two models.)  Moreover, with $S_{1400}<100$\,$\mu$Jy,
PSR~J0646$-$5455 is one of the radio-faintest known $\gamma$-ray pulsars, though its flux lies above the 30\,$\mu$Jy ``radio-faint'' threshold adopted by the LAT team \citep{Abdo13_2PC,Smith23}.  Although radio and $\gamma$-ray fluxes are not necessarily correlated \citep{Smith23}, this low radio flux is also suggestive of a greater distance.  Confirming
such a low efficiency would provide a valuable constraint on models of
$\gamma$-ray emission \citep[see][]{Philippov22}, motivating a future
parallax measurement.  (The precision of parallax measurements with LAT pulsar timing is generally too low to be useful.)

PSR~J0646$-$5455 has a canonical Vela-like pulse profile, with two
caustic-shaped $\gamma$-ray peaks separated by about $\Delta=0.4$ and
trailing a narrow radio peak by about $\delta=0.1$ (see 3PC), further indicating that this configuration is common across orders of magnitude in $\dot{E}$, light cylinder size, and magnetic field strength.

The $\gamma$-ray brightness, along with a 2.5\,ms spin period and narrow
pulse profile, make this pulsar suitable for high-precision timing.
Following the methods of \citet{Ajello22}, we obtained a single-pulsar
gravitational wave background upper limit (95\% confidence) of $\mathcal{A}_{\mathrm{gwb}}<1.3\times10^{-13}$, making J0646$-$5455 one of the 25 best-timed pulsars in the GPTA.

\subsection{PSR~J1803$-$4817}

% I plotted Pb vs P0 from psrcat, nothing remarkable about a 3.6ms pulsar
% in a 100-day orbit.
PSR~J1803$-$4817 is in the most eccentric binary of the four, and like the other two non-eclipsing systems, the residual eccentricity ($e = 0.9\times10^{-4}$) is roughly consistent with the predictions from tidal circularization due to coupling
with convective cells in the companion envelope \citep{Phinney92}, $\epsilon \sim 1.5\times10^{-4}(P_b/100\,\mathrm{d}) \approx 1.4\times10^{-4}$.  Though
there is a modest correlation between spin period and orbital period observed in cataloged MSPs, the
3.7\,ms spin period is not unusually fast compared to similarly-wide binaries in
the Galactic field.

%The 90\,d
%Ellipticity of J1803 orbit, evolution implications.  (How do you get a 3ms pulsar and a 90d orbit?)
%--> NB, the eccentricity is more or less in line with the predictions of Phinney 1992!  what a guy
%Unusually soft gamma-ray spectrum.

\subsection{PSR~J1833$-$3840}
PSR~J1833$-$3840 is the only eclipsing pulsar in the sample, and with a
minimum companion mass $<$0.01\msun, it is a black widow.  It has an
unusually steep spectrum, with an index of nearly $-3$, which may be a hallmark
of some high-$\dot{E}$ MSPs \citep[e.g.][]{Backer82,Fruchter88}.  Indeed, \citet{Frail18} independently noted this source as a good MSP candidate due to its steep spectrum.

J1833$-$3840 is also a high-$\dot{E}$ pulsar, and although the $\gamma$-ray
light curve suffers from a poor S/N, it is clear that the $\gamma$-ray
emission is largely aligned with the radio peaks, which is also the case
for both the original MSP and the original black widow pulsar
\citep{Guillemot12a}.  It has been suggested \citep[e.g.][]{Bangale24} that spider pulsars have
faster periods than the isolated and/or less compact systems; with a period
of only 1.87\,ms, this pulsar provides further support.  This low spin period, along with a relatively low surface
magnetic field, invite comparison to PSR~J0952$-$0607, a soft radio
spectrum, fast-spinning black widow.  Modeling of the optical light
curve of the heated companion of PSR~J0952$-$0607 suggests a heavy neutron star
\citep{Romani22}, motivating optical follow-up of J1833$-$3840.

Finally, J1833$-$3840 has the longest period of any known black widow system and thus gives insight into the evolution of compact, eclipsing binaries.  Studies differ as to whether black widows and redbacks (with similar orbital period but companion masses of order 0.1\msun) are discrete populations \citep{Chen13} or consecutive evolutionary phases, with redbacks evolving into black widows \citep{Misra24}.  In the latter scenario, forming a binary with $P_b=0.9\,d$ would require extremely efficient evaporation of the companion.  Characterization of the companion elemental abundance and other properties strengthen constraints on the binary evolution, further motivating optical studies.

%The velocity for this pulsar is also 160pm50 km/s, but with the substantial
%distance uncertainty, not as interesting.

\subsection{PSR~J2045$-$6837}

With a proper motion of $25\pm2$\,mas\,yr$^{-1}$,the implied spatial
velocity at the DM distance of 1.3\,kpc is 152$\pm$14\,km\,s$^{-1}$, about
twice the mean two-dimensional speed observed for recycled pulsars
\citep{Hobbs04}.  The $\gamma$-ray efficiency, about 4\%, is also
relatively low.  These anomalies are modest, but can be alleviated if
the true distance to the pulsar is $<$1\,kpc, as is the case for the NE2001 model distance.  Alternatively, if the high
spatial velocity is confirmed, it would provide a useful datum for the strength of supernova kicks relative to requirements for binary disruption.

PSR~J2045$-$6837 has a steeper-than-average radio spectrum, and the
leading edge of its pulse profile is highly polarized (Figure
\ref{fig:j2045_pol}), enabling the only measurement of RM for the
sample.

\subsection{Suitability for Radio PTAs}
\label{sec:radio_ptas}
We give a rough estimate of the typical radio timing precision for these
pulsars with the \textit{Murriyang} UWL in Table \ref{tab:timing} but
provide more context here.  These measurements are based only on the
fold-mode data, which is limited.  In particular, some of the brightest
epochs for PSR~J0646$-$5455 were observed only in ``search-mode'', so that
the reported timing precision is conservative.  PSR~J1833$-$3840 could
reliably deliver timing precision at the 2--3$\mu$s level (in one hour)
and, due to the larger DM, is less subject to scintillation.  The other
pulsars are too faint to time with high precision without dynamic
scheduling to target scintillation maxima.  On the other hand, given the
relatively steep spectra for all pulsars save PSR~J0646$-$5455---and
especially that of the eclipsing J1833$-$3840---MeerKAT and the future SKA
(both Mid and Low) should regularly achieve $\mu$s-precision measurements
for these pulsars, making them plausible PTA candidates.

\subsection{The Survey in Retrospect}
\label{sec:retro}

\begin{deluxetable}{lll}
\footnotesize
\tablecaption{Pulsars Discovered in Survey Targets\label{tab:disco}}
\tablehead{
\colhead{Target} &
\colhead{Pulsar}  &
\colhead{Telescope/Survey}}
\startdata
J0312$-$0920 & J0312$-$0921 & GBT \\ %Tabassum
J0744$-$2435 & J0744$-$2525$^\dagger$ & E@H \\
J0838$-$2827 & J0838$-$2827 & TRAPUM$^{\mathrm{T}24}$ \\
J0952$-$0608 & J0952$-$0607 & LOFAR$^{\mathrm{B}17}$ \\
J0646$-$5455 & J0646$-$5455 & this work \\
J1204$-$5033 & J1207$-$5050 & GMRT$^{\mathrm{B}21}$ \\%(previous discovery, oops)
J1208$-$6238 & J1208$-$6238$^\dagger$ & E@H \\
J1259$-$8149 & J1259$-$8148 & TRAPUM \\ %no ref?
J1304+1203 & J1304+12 & AO \\ % Cromartie?
J1335$-$5656 & J1335$-$5656 & E@H \\ % FGRP5
J1345$-$2612 & J1346$-$2610 & TRAPUM \\ % no ref?
%J1526$-$3811 & J1529$-$3828 & HTRU \\ % need to confirm
%J1533$-$5232 3 WVU possibly J1537$-$5312? HTRU
J1544$-$2534 & J1544$-$2555 & TRAPUM \\ % no ref
J1555$-$2908 & J1555$-$2908 & GBT$^{\mathrm{R}22}$ \\
J1611$-$6011 & J1603$-$6011 & TRAPUM \\ % no ref
J1623$-$6935 & J1623$-$6936 & TRAPUM$^{\mathrm{C}23}$ \\
J1649$-$3010 & J1649$-$3012 & E@H \\ % FGRP5
J1653$-$0158 & J1653$-$0158 & E@H$^{\mathrm{N}20}$ \\
J1802$-$4718 & J1803$-$4719 & this work \\
J1823$-$3543 & J1823$-$3543 & TRAPUM$^{\mathrm{C}23}$ \\
J1832$-$3840 & J1833$-$3840 & this work\\
J1858$-$5425 & J1858$-$5422 & TRAPUM$^{\mathrm{C}23}$ \\
J2029$-$4237 & J2029$-$4239 & TRAPUM \\ % no ref
J2045$-$6835 & J2045$-$6837 & this work \\
\enddata
\tablecomments{
Now-known MSPs and young pulsars$^\dagger$ associated with the
pulsar search targets.  Information about unpublished pulsars is available
  from the Galactic MSP
  list at \url{https://www.astro.umd.edu/~eferrara/pulsars/GalacticMSPs.txt} and the Einstein@Home website, \url{https://einsteinathome.org/gammaraypulsar/FGRP1_discoveries.html}.
  References:
(B17) \citet{Bassa17}
(N20) \citet{Nieder20}
(B21) \citet{Bhattacharyya21}
(R22) \citet{Ray22}
(C23) \citet{Clark23a}
  (T24) \citet{Thongmeearkom24}.}
\vspace{-0.8cm}
\end{deluxetable}

%\citet{Ransom11}: 3/25 = 12\%
%\citet{Bangale24}: Bangale: 18/49 = 37\%
%\citet{Keith11}: 2/11 = 18\%
%\citet{Kerr12}: 5/14 = 36\%
%\citet{Camilo15}: 11/56 = 20\%, or 6/42=14\%.
%\citet{Barr13}: 1/289
%\citet{Clark23a}: 9/79 = 11\%
Targeting pulsar-like LAT sources has yielded good return
on telescope time investment.  Early surveys with the GBT at 820\,MHz
\citep{Ransom11} and 350\,MHz \citep{Bangale24} had a purity of
3/25=12\% and 18/49=37\%, respectively\footnote{Because it was
published well after the survey completion, \citet{Bangale24} provided
a careful analysis of subsequent discoveries in their target list, as
we have here.  For other surveys, only the initially reported number
of MSP discoveries is considered.}.  Meanwhile, early surveys with the
Parkes observatory had 2/11=18\% \citep{Keith11}, 5/14=36\%
\citep{Kerr12} purity.  After these initial successes, there was some
indication of diminishing returns.  \citet{Camilo15} reported the
detection of 11/56=20\% new pulsars, but these included the
discoveries of \citet{Kerr12}; the new 2FGL targets only yielded 6/42=14\%
purity.  Using the Effelsberg telescope, \citet{Barr13} searched 289
LAT targets based on 1FGL and found only 1 MSP.  In this work, we report only 4
discoveries from 80 4FGL-based targets, for a comparatively low 5\% success rate.

However, in the years since these original discoveries were made, the
sources have been searched with other telescopes, and 17 more MSPs and 2 young pulsars have
been found, as listed in Table \ref{tab:disco}.  Five of these are
radio-quiet pulsars discovered in direct searches of the LAT data with Einstein@Home \citep[e.g.][]{Clark18}.  A further two targets are likely to be pulsars: J0523$-$2527 is a high-confidence compact binary (redback) candidate \citep{Strader14} while J1120$-$2214 is an optical NS-proto WD system that also likely hosts a pulsar \citep{Swihart22}.

For the radio-loud pulsars, a common feature for successful detection is the use of a more sensitive or
lower-frequency telescope.  And indeed, the pulsars discovered in this
survey are generally not very radio bright, with PSR~J0646$-$5455 having a
mean flux density at 1.4\,GHz well below 0.1\,mJy; this is a typical
sensitivity threshold for a radio survey.  The remaining three pulsars all have
relatively steep spectra and are indicative of the pulsars well-suited for low-frequency observations.  Finally, repeated observations also play an important role.  Because it is observable at a favorable local sidereal time range and because of its quality as a pulsar candidate, we observed J0838$-$2827 no fewer than 11 times, with no detections.  The TRAPUM team were able to discover pulsations with MeerKAT \citep{Thongmeearkom24}, but nonetheless required many additional attempts to verify the discovery detection of this extremely enshrouded redback.

When considering these subsequent pulsar detections, the
target selection increases to 29\% purity, indicating that relatively
faint, new sources in 4FGL continue to provide excellent targets for radio
telescopes.

We can draw a few more conclusions from this analysis.  First, the surveys which used manually-curated target lists (\citet{Bangale24}, \citet{Kerr12}, and this work) tend to have the highest purity, although classification with
earlier machine learning techniques \citep[e.g.][]{sazParkinson16} has
also been successful.  Conversely, \citet{Barr13} cast a very wide net,
observing nearly half of the 600 unidentified sources in the 1FGL catalog,
of which the majority are likely blazars.  Moreover, many of the targets of \citet{Barr13} were in the Galactic plane.  While this is a natural choice when looking for pulsars, the analysis techniques for $\gamma$-ray sources at low Galactic latitude have advanced substantially since 1FGL, and \citet{Barr13} noted that many of their targets were absent or had moved more than the telescope beam size when compared to 2FGL.  While Galactic $\gamma$-ray source positions are likely now more reliable, it remains true that targeting higher Galactic latitudes also targets the relatively nearby MSP population to which LAT is most sensitive.

On the other hand, it may be that accepting lower survey efficiency will be necessary for improved understanding of the $\gamma$-ray MSP population. The new MSPs reported here all have $\gamma$-ray efficiencies
well below the mean reported in 3PC.  It is possible that these
efficiencies are in error (e.g. all of the DM distance estimates are too
low).  But it is also possible that the discoveries of these pulsars are beginning to
correct a bias in the LAT population, which was established largely through
search efforts like the one described here.  Those are naturally biased
towards the brightest sources, which can be bright both by being close and
by being particularly efficient, either through favorable beaming or a
truly high $\gamma$-ray luminosity.  Quantifying the full spread of the
range of MSP efficiencies is key both to understanding their magnetospheres
and the production of $\gamma $rays \citep[e.g.][]{Philippov22} and to
determining the contribution of MSP populations to features like the Fermi
Galactic center excess \citep[e.g.][]{Ackermann17_gce}.

\begin{acknowledgements}
We thank Lawrence Toomey for assistance in wrangling the data and Jane
Kaczmarek for help in scheduling filler observations which helped in
obtaining phase connection.

The \textit{Fermi} LAT Collaboration acknowledges generous ongoing support
  from a number of agencies and institutes that have supported both the
  development and the operation of the LAT as well as scientific data
  analysis.  These include the National Aeronautics and Space
  Administration and the Department of Energy in the United States, the
  Commissariat \`a l'Energie Atomique and the Centre National de la
  Recherche Scientifique / Institut National de Physique Nucl\'eaire et de
  Physique des Particules in France, the Agenzia Spaziale Italiana and the
  Istituto Nazionale di Fisica Nucleare in Italy, the Ministry of
  Education, Culture, Sports, Science and Technology (MEXT), High Energy
  Accelerator Research Organization (KEK) and Japan Aerospace Exploration
  Agency (JAXA) in Japan, and the K.~A.~Wallenberg Foundation, the Swedish
  Research Council and the Swedish National Space Board in Sweden.
 
Additional support for science analysis during the operations phase is
  gratefully acknowledged from the Istituto Nazionale di Astrofisica in
  Italy and the Centre National d'\'Etudes Spatiales in France. This work
  performed in part under DOE Contract DE-AC02-76SF00515.

Murriyang, the Parkes radio telescope, is part of the Australia Telescope
  National Facility, which is funded by the Australian Government for
  operation as a National Facility managed by CSIRO.

Work at NRL is supported by NASA, in part by Fermi Guest Investigator grant
  NNG22OB35A.

The MeerKAT telescope is operated by the South African Radio Astronomy Observatory, which is a facility of the National Research Foundation, an agency of the Department of Science and Innovation.

Observations used the FBFUSE and APSUSE computing clusters for data acquisition, storage and analysis. These clusters were funded, designed and installed by the Max-Planck-Institut-für-Radioastronomie (MPIfR) and the Max-Planck-Gesellschaft. FBFUSE perforns beamforming operations in real-time using the \texttt{mosaic}\footnote{\url{https://github.com/wchenastro/Mosaic}} software stack \citep{Chen2021+FBFUSE}.

\end{acknowledgements}

\facilities{Fermi}

\bibliographystyle{aasjournal}
%\bibliographystyle{apj_hyperref}
%\bibliography{three,sr}

\input{three.bbl}
\appendix

\section{Targets}
\label{sec:targets}

Table \ref{tab:targets} gives the full record of pointing positions and integration lengths for Parkes observations made for P814.  The ``Target'' column indicates the name by which data can be found in the CSIRO online archive.  For the detected pulsars, we include the search epochs but exclude subsequent timing observations.  We also exclude some observations logged with the P814 project code but which were not intended for pulsar searches, e.g. timing of the newly-discovered PSR~J1208$-$6238.  NB that this table does not include observations of these targets carried out under other search campaigns, potentially at other telescopes, under the auspices of the Pulsar Search Consortium\footnote{A substantial but incomplete set of logs for PSC observations is available upon request to the authors.} \citep{Ray12} or otherwise.  (See Table \ref{tab:disco} for the campaigns which have subsequently discovered pulsations in some of these targets.)

We have compared the pointing positions to the 4FGL-DR4 source catalog and indicate the distance to the nearest source.  In most cases, the separation is much less than the size of the telescope beam, about 7\arcmin{} radius.  Because the target list was based on a preliminary version of the catalog, in a few cases, the targeted source is not in the 4FGL catalog, and when no catalog source exists within 0.3\arcdeg{}, no counterpart is listed.  Due to a transcription error, an incorrect pointing position was used for (now) PSR~J1555$-$2908, which has a 4FGL counterpart.

\input{obslog.tex}

\end{document}

%% file: timing_table.tex
\tabletypesize{\scriptsize}
\begin{deluxetable*}{lcccc}
\tablewidth{0pt}
\tablecaption{Timing and $\gamma$-ray Spectral Parameters for the
Discovered MSPs\label{msp_timing1}}
%--------------------------------------------------------------------------------------------------------------------------------------------------------------------------------------------------------------------------------
\tablehead{
\colhead{Timing Parameters} &
\colhead{PSR J0646$-$5455}  &
\colhead{PSR J1803$-$4719}  &
\colhead{PSR J1833$-$3840}  &
\colhead{PSR J2045$-$6837} }
\startdata
Fermi-LAT 4FGL Source \dotfill & 
J0646.4$-$5455    &
J1802.8$-$4719    &
J1833.0$-$3840    &
J2045.9$-$6835  \\
%--------------------------------------------------------------------------------------------------------------------------------------------------------------------------------------------------------------------------------
Right Ascension (J2000)  \dotfill   &
$06 \hr 46 \mi 14 \fs 62993(7)$ &
$18 \hr 03 \mi 00 \fs 974(3)$ &
$18 \hr 33 \mi 04 \fs 58231(9)$ &
$20 \hr 45 \mi 40 \fs 0623(2)$ \\
Declination     (J2000)  \dotfill   &
$-54 \arcdeg 55 \arcmin14 \farcs 103(1)$ &
$-47 \arcdeg 19 \arcmin08 \farcs 319(9)$ &
$-38 \arcdeg 40 \arcmin46 \farcs 072(5)$ &
$-68 \arcdeg 37 \arcmin26 \farcs 158(1)$ \\
Pulsar Period (ms)  \dotfill    &
%  2.50728557771931(5) &
%  3.6693711493223(8) &
%  1.8660672158606(3) &
%  2.9622548584399(2) \\
  2.5073 &
  3.6694 &
  1.8661 &
  2.9623 \\
Period Derivative, {\it \.{P}} (10$^{-20}$ s\,s$^{-1}$) \dotfill &
0.18122(8) &
3.437(1) &
1.7770(4) &
1.2802(9) \\
Period Reference Epoch (MJD)   \dotfill   &
57523.000 &
%57419.9735192558 &
%57549.348968 &
%60263.6052712129 \\
57419.974 &
57549.349 &
60263.605 \\
Dispersion Measure (pc\,cm$^{-3}$) \dotfill    &
40.2289(6) &
41.642(2)  &
78.6606(9) &
21.0717(6) \\
Proper motion in RA $\mu_\alpha\cos\delta$ (mas\,yr$^{-1}$) \dotfill &
  4.3(2)     &
  -1.5(5)     &
  -5(2)     &
  16(2)    \\
Proper motion in Dec $\mu_\delta$  (mas\,yr$^{-1}$) \dotfill   &
  2.2(2) &
  -3(2) &
  -5(3) &
  -20(3) \\
Position Reference Epoch (MJD)   \dotfill   &
60267.86 &
60222.95 &
57549.35 &
60379.90 \\
Orbital Period (days)       \dotfill                    &
9.618521110(3) &
90.44184836(9) &
0.900452049(8) &
5.71727490(1) \\
Projected Semi-Major Axis (lt-s)    \dotfill                  &
6.551460(2) &
34.430149(3) &
0.061478(3) &
3.978860(2) \\
Orbital Eccentricity (10$^{-5}$)       \dotfill                    & 
3.50(3) &
8.96(2) &
10(8)  &
1.37(10) \\
Epoch of Ascending Node (MJD)     \dotfill                  &
60271.1388002(3) &
57419.993662(2) &
57311.052610(6) &
60263.6049804(7) \\
%---------------------------------------------------------------------------------------------------------------------------------------------------------------------------------------------------------------------------------
\cutinhead{Timing-Derived Parameters} 
Mass Function ($10^{-3}$ \msun)   \dotfill                  &
3.26  &
5.36  &
3.08$\times10^{-4}$ &
2.07 \\
Minimum Companion Mass (\msun)    \dotfill                  &
$\geq$\,0.198    &
$\geq$\,0.238    &
$\geq$\,0.00828  &
$\geq$\,0.168  \\
%Galactic Longitude (deg)    \dotfill                      & 111.3
%& 124.7       & 131.7       & 153.7       & 284.1              \\
%Galactic Latitude (deg)     \dotfill                      & $-$52.8
%& $-$14.2     & 14.2        & $-$11.1     & 22.8               \\
%DM-derived Distance (NE2001, kpc)    \dotfill              & 0.7
%& 2.3        & 0.6       & 1.7     & 1.7                \\
DM-derived Distance (YMW16, kpc)    \dotfill              &
0.37 &
1.2 &
4.7 &
1.3 \\
DM-derived Distance (NE2001, kpc)    \dotfill              &
1.8 &
1.2 &
2.1 &
0.82 \\
%Surface Mag. Field, $B$ ($10^8$\,G)   \dotfill                & 1.8
%& 1.8       & 2.3         & 1.5     & 1.2                \\
%Characteristic Age (Gyr)    \dotfill                      & 4.2
%& 4.1       & 2.9       & 7.4     & 6.3                \\
Spin-down Luminosity, $\dot{E}$ ($10^{34}$\,erg\,s$^{-1}$)  \dotfill &
0.4539 &
2.746 &
10.80 &
1.944 \\
%----------------------------------------------------------------------------------------------------------------------------------------------------------------------------------------------------------------------------
Transverse velocity (km~s$^{-1}$) \dotfill                          &
8.4(4) &
19(8) &
160(46) &
153(14) \\
%Acceleration $\perp$ to plane (10$^{-14}$ s s$^{-1}$) \dotfill
%& $-$5.8(4)   & $-$4.9(3)       & 2.1(3)            & $-$4.0(2)     &
%4.4(2)    \\
%Acceleration $\parallel$ to plane  (10$^{-14}$ s s$^{-1}$) \dotfill
%& $-$0.7(1)   & $-$0.73(2)      & $-$0.071(8)     & 2.1(4)      &
%$-$2.0(4) \\
Corrected {\it \.{P}} (10$^{-20}$ s s$^{-1}$) \dotfill &
0.190(4) &
3.40(1) &
1.63(6) &
0.7(2) \\
Corrected $\dot{E}$ ($10^{34}$\,erg\,s$^{-1}$)  \dotfill &
0.48(1) &
2.721(8) &
9.9(3) &
1.1(3) \\
%Corrected $B$ ($10^8$\,G)   \dotfill
%& 1.7       & 1.8           & 2.3           & 1.5       & 1.2       \\ 
%---------------------------------------------------------------------------------------------------------------------------------------------------------------------------------------------------------------------------------
\cutinhead{Radio Emission Properties}
Rotation Measure (RM) (rad m$^{-2}$)  \dotfill            &
\dots   &
\dots   &
\dots   &
17(3) \\
Flux Density at 1010\,MHz (mJy)          \dotfill              &
0.11(5) &
0.5(2) &
1.00(2) &
0.5(3) \\
Flux Density at 1730\,MHz (mJy)          \dotfill              &
0.09(6) &
0.2(1) &
0.21(2) &
0.15(11) \\
Flux Density at 3260\,MHz (mJy)         \dotfill              &
0.04(2) &
0.06(8) &
0.03(1) &
0.04(1) \\
Flux Density at 1400\,MHz (mJy), $S_{1400}$         \dotfill              &
0.09(3) &
0.3(1) &
0.38(2) &
0.25(11) \\
Spectral Index, $\alpha$   \dotfill              &
-0.8(6) &
-1.9(10) &
-2.9(1) &
-2.3(6) \\
TOA Precision at 1010\,MHz ($\mu$s)  \dotfill &
13 & 6.4 & 2.5 & 4.3 \\
TOA Precision at 1730\,MHz ($\mu$s)  \dotfill &
5.5 & 9.9 & 3.0 & 15 \\
TOA Precision at 3260\,MHz ($\mu$s)  \dotfill &
6.8 & 18 & 25 & 27 \\
%---------------------------------------------------------------------------------------------------------------------------------------------------------------------------------------------------------------------------------
\cutinhead{$\gamma$-ray Spectral Properties}
%K ($10^{-12}$ cm$^{-2}$\,s$^{-1}$\,MeV$^{-1}$)   \dotfill       &
%3.0(5)     & 2.4(2)     & 3.6(6)  & 2.9(2)  & 1.8(2)      \\
%Photon Index $\Gamma$          \dotfill                   &
%1.7(2) &
%2.6(7) &
%2.1(2) &
%1.7(2) \\
Peak Energy, E$_{\mathrm{p}}$ (GeV)         \dotfill                   & 
2.0(2) &
0.5(4) &
1.3(4) &
2.2(3) \\
Spectral Curvature, d$_{\mathrm{p}}$ \dotfill &
1.3(3) &
0.3(3) &
0.6(3) &
1.2(3) \\
Photon Index at 100\,MeV, $\Gamma_{100}$       \dotfill                   &
0.3(4) &
1.7(4) &
1.2(4) &
0.4(4) \\
Energy Flux, G$_{100}$ ($10^{-12}$ erg\,cm$^{-2}$\,s$^{-1}$)  \dotfill       &
%2.74(27)
2.7(3) &
%3.05(71)
3.1(7) &
%2.71(45)
2.7(4) &
%2.37(26)
2.4(3) \\
Efficiency                           \dotfill  &
% preliminary computation
0.009(4) &
0.019(9) &
0.07(3) &
0.04(2) \\
\enddata
\tablecomments{Numbers in parentheses represent 1-$\sigma$
uncertainties in the last digit as determined from the timing
pipeline described in the main text.  For the sake of legibility, we truncate the precision of some parameters: exact values are available in the online ephemerides.
Minimum companion masses were calculated assuming a pulsar mass of
1.4\,\msun. The gamma-ray spectral parameters are from 4FGL-DR4
\citep{4FGL_DR4} using a PLEC4 model with $b$=0.67.  We have converted these parameters to the more physical quantities identified in 3PC, namely the energy at which the spectral energy density peaks, $E_{\mathrm{p}}$, the logarithmic curvature $d_{\mathrm{p}}$ evaluated at $E_{\mathrm{p}}$; and the logarithmic slope at 100\,MeV, $\Gamma_{100}$.  
 $\gamma$-ray efficiency is estimated as $4\pi G_{100} d^2/\dot{E}$; this simple formulation assumes that the $\gamma$-ray beams illuminate the sky uniformly. All distance-dependent quantities are calculated using the YMW16 \citep{YMW17} electron density model and, where needed, assuming 30\% uncertainty.  NE2001 \citep{Cordes02} model distances are reported for comparison.
   The flux density measurements are reported as the mean and standard deviation of all data, while the TOA precision estimates are based only on the limited subset of fold-mode data (see Figure \ref{fig:fluxes}).}
\label{tab:timing}
\end{deluxetable*}

%% file: obslog.tex
\tabletypesize{\scriptsize}
\startlongtable
\begin{deluxetable*}{clrrrcc}
\centerwidetable
\movetableright=3cm
\tablewidth{0pt}
\tablecaption{Pointing Information for Sources Searched\label{msp_search}}
%--------------------------------------------------------------------------------------------------------------------------------------------------------------------------------------------------------------------------------
\tablehead{
\colhead{Target} &
\colhead{Date}  &
\colhead{Length (s)}  &
\colhead{R.A.}  &
\colhead{Decl.}  &
\colhead{4FGL-DR4} &
\colhead{Separation (\arcdeg)}}
\startdata
\hline
J0245$-$5951 & 2016-11-04 & 3606 & 41.37083 & -59.85000 & J0245.4$-$5950 & 0.018 \\
\hline
J0312$-$0920 & 2016-11-04 & 3606 & 48.04583 & -9.33806 & J0312.1$-$0921 & 0.024 \\
 & 2016-11-06 & 2393 &  &  &  &  \\
 & 2017-02-13 & 968 &  &  &  &  \\
 & 2017-02-13 & 3606 &  &  &  &  \\
\hline
J0414$-$4259 & 2015-10-17 & 3606 & 63.70000 & -42.98694 & J0414.7$-$4300 & 0.025 \\
 & 2015-10-17 & 4006 &  &  &  &  \\
 & 2015-11-22 & 3606 &  &  &  &  \\
 & 2015-11-22 & 5627 &  &  &  &  \\
 & 2016-01-17 & 3606 &  &  &  &  \\
\hline
J0443$-$6651 & 2016-11-04 & 3606 & 70.88333 & -66.85556 & J0443.3$-$6652 & 0.018 \\
\hline
J0523$-$2527 & 2016-11-04 & 3606 & 80.82917 & -25.45528 & J0523.3$-$2527 & 0.008 \\
\hline
J0628+0519 & 2017-02-21 & 3606 & 97.18333 & 5.32139 &  &  \\
 & 2017-02-22 & 2706 &  &  &  &  \\
\hline
J0646$-$5455 & 2016-02-03 & 3606 & 101.60833 & -54.91889 & J0646.4$-$5455 & 0.010 \\
 & 2016-03-03 & 1313 &  &  &  &  \\
 & 2016-03-28 & 3606 &  &  &  &  \\
 & 2016-03-28 & 1803 &  &  &  &  \\
 & 2016-03-28 & 3606 &  &  &  &  \\
 %& 2016-04-30 & 3606 &  &  &  &  \\
 %& 2016-06-16 & 3606 &  &  &  &  \\
 %& 2017-02-16 & 3606 &  &  &  &  \\
 %& 2017-02-17 & 3606 &  &  &  &  \\
 %& 2017-02-20 & 3606 &  &  &  &  \\
 %& 2017-02-21 & 3606 &  &  &  &  \\
 %& 2017-02-22 & 3606 &  &  &  &  \\
 %& 2017-03-19 & 3606 &  &  &  &  \\
\hline
J0744$-$2535 & 2017-02-16 & 3606 & 116.04583 & -25.39944 & J0744.0$-$2525 & 0.043 \\
 & 2017-02-17 & 3606 &  &  &  &  \\
 & 2017-02-20 & 3606 &  &  &  &  \\
 & 2017-03-19 & 3606 &  &  &  &  \\
\hline
J0745$-$4028 & 2017-02-21 & 3606 & 116.25000 & -40.47111 & J0744.9$-$4028 & 0.004 \\
\hline
J0749+1325 & 2016-11-05 & 3597 & 117.41250 & 13.43306 & J0749.6+1324 & 0.026 \\
\hline
J0754$-$3953 & 2015-10-17 & 3606 & 118.70833 & -39.88694 & J0754.9$-$3953 & 0.026 \\
 & 2015-11-22 & 3606 &  &  &  &  \\
 & 2015-11-22 & 3606 &  &  &  &  \\
 & 2016-02-03 & 3606 &  &  &  &  \\
 & 2016-03-28 & 3606 &  &  &  &  \\
 & 2016-04-30 & 3606 &  &  &  &  \\
 & 2016-06-17 & 3606 &  &  &  &  \\
\hline
J0758$-$1451 & 2016-11-04 & 3606 & 119.70000 & -14.85278 & J0758.8$-$1450 & 0.016 \\
\hline
J0816$-$0007 & 2016-11-05 & 3616 & 124.11667 & -0.12417 & J0816.4$-$0007 & 0.006 \\
\hline
J0826$-$5054 & 2016-11-04 & 3606 & 126.53333 & -50.91028 & J0826.1$-$5053 & 0.012 \\
 & 2017-02-20 & 3606 &  &  &  &  \\
\hline
J0838$-$2827 & 2016-11-05 & 3606 & 129.69583 & -28.45667 & J0838.7$-$2827 & 0.005 \\
 & 2017-02-16 & 3606 &  &  &  &  \\
 & 2017-02-16 & 3606 &  &  &  &  \\
 & 2017-02-16 & 3606 &  &  &  &  \\
 & 2017-02-16 & 3606 &  &  &  &  \\
 & 2017-02-16 & 3606 &  &  &  &  \\
 & 2017-02-17 & 3606 &  &  &  &  \\
 & 2017-02-17 & 3606 &  &  &  &  \\
 & 2017-02-20 & 3606 &  &  &  &  \\
 & 2017-02-21 & 3606 &  &  &  &  \\
 & 2017-02-22 & 3606 &  &  &  &  \\
\hline
J0919$-$6204 & 2015-10-17 & 3606 & 139.93333 & -62.07778 & J0919.5$-$6203 & 0.027 \\
 & 2015-10-17 & 3606 &  &  &  &  \\
 & 2015-10-17 & 2993 &  &  &  &  \\
 & 2015-11-22 & 3606 &  &  &  &  \\
\hline
J0922$-$6316 & 2016-02-03 & 3606 & 140.55833 & -63.26861 & J0921.7$-$6317 & 0.055 \\
 & 2016-03-28 & 3606 &  &  &  &  \\
 & 2016-04-30 & 1083 &  &  &  &  \\
 & 2016-06-18 & 3606 &  &  &  &  \\
 & 2017-02-17 & 3606 &  &  &  &  \\
\hline
J0933$-$6233 & 2016-02-03 & 3606 & 143.49167 & -62.55028 & J0933.8$-$6232 & 0.008 \\
 & 2016-03-28 & 3606 &  &  &  &  \\
 & 2016-06-18 & 3606 &  &  &  &  \\
\hline
J0952$-$0608 & 2016-11-04 & 3606 & 148.05417 & -6.13333 & J0952.1$-$0607 & 0.008 \\
\hline
J0953$-$1509 & 2016-11-05 & 3606 & 148.40417 & -15.15639 & J0953.6$-$1509 & 0.002 \\
\hline
J1107$-$7724 & 2016-11-04 & 3606 & 166.83750 & -77.41639 &  &  \\
\hline
J1117$-$4840 & 2016-02-03 & 3606 & 169.27500 & -48.67694 & J1117.5$-$4839 & 0.076 \\
\hline
J1120$-$2204 & 2016-11-04 & 3606 & 170.00000 & -22.07444 & J1120.0$-$2204 & 0.004 \\
\hline
J1126$-$5006 & 2017-02-16 & 3606 & 171.51250 & -50.10833 & J1126.0$-$5007 & 0.011 \\
 & 2017-02-21 & 3606 &  &  &  &  \\
\hline
J1145$-$6822 & 2016-11-04 & 3606 & 176.42917 & -68.37750 & J1146.0$-$6822 & 0.033 \\
\hline
J1204$-$5033 & 2015-10-17 & 3606 & 181.16667 & -50.55806 & J1204.5$-$5032 & 0.017 \\
 & 2015-10-17 & 3606 &  &  &  &  \\
 & 2015-11-22 & 3606 &  &  &  &  \\
 & 2015-11-22 & 3606 &  &  &  &  \\
 & 2015-11-22 & 1747 &  &  &  &  \\
\hline
J1207$-$4536 & 2017-02-20 & 3606 & 181.88333 & -45.60028 & J1207.4$-$4536 & 0.014 \\
\hline
J1207$-$4537 & 2016-02-03 & 3606 & 181.87083 & -45.61833 & J1207.4$-$4536 & 0.006 \\
\hline
J1208$-$5512 & 2016-11-04 & 3606 & 182.12917 & -55.20694 & J1208.5$-$5512 & 0.009 \\
\hline
%J1208$-$6238 & 2016-03-28 & 9006 & 182.05833 & -62.63389 & J1208.2$-$6237 & 0.003 \\
% & 2016-04-14 & 15376 &  &  &  &  \\
% & 2016-06-18 & 5333 &  &  &  &  \\
\hline
J1236+1129 & 2017-02-17 & 3606 & 189.19583 & 11.48417 & J1235.9+1136 & 0.240 \\
\hline
J1259$-$8149 & 2016-11-04 & 3606 & 194.85833 & -81.82611 & J1259.0$-$8148 & 0.017 \\
 & 2017-02-20 & 2773 &  &  &  &  \\
\hline
J1304+1203 & 2016-11-05 & 3606 & 196.12083 & 12.05833 & J1304.4+1203 & 0.006 \\
\hline
J1335$-$5656 & 2016-02-03 & 3606 & 203.77500 & -56.93611 & J1335.0$-$5656 & 0.012 \\
 & 2017-02-21 & 3606 &  &  &  &  \\
\hline
J1345$-$2612 & 2016-11-05 & 3606 & 206.49167 & -26.20778 & J1345.9$-$2612 & 0.010 \\
\hline
J1404$-$5236 & 2016-11-05 & 3606 & 211.22917 & -52.60278 & J1404.8$-$5237 & 0.028 \\
\hline
J1416$-$5021 & 2017-02-16 & 3606 & 214.19583 & -50.35833 & J1416.7$-$5023 & 0.033 \\
 & 2017-02-22 & 3606 &  &  &  &  \\
\hline
J1435$-$3906 & 2016-02-03 & 3606 & 218.98750 & -39.10750 &  &  \\
\hline
J1439$-$5142 & 2016-02-03 & 3606 & 219.78750 & -51.70556 & J1439.2$-$5142 & 0.012 \\
\hline
J1512$-$7132 & 2015-10-17 & 698 & 228.24583 & -71.53417 & J1513.2$-$7131 & 0.026 \\
 & 2015-10-17 & 2128 &  &  &  &  \\
\hline
J1517$-$4446 & 2016-11-05 & 1233 & 229.37500 & -44.76889 & J1517.7$-$4446 & 0.038 \\
\hline
J1526$-$3811 & 2016-02-03 & 3606 & 231.64167 & -38.18417 & J1526.6$-$3810 & 0.021 \\
\hline
J1533$-$5232 & 2017-02-16 & 3606 & 233.47917 & -52.54944 & J1534.0$-$5232 & 0.013 \\
 & 2017-02-22 & 3606 &  &  &  &  \\
 & 2017-03-19 & 3606 &  &  &  &  \\
\hline
J1544$-$2554 & 2016-11-05 & 3606 & 236.05833 & -25.91167 & J1544.2$-$2554 & 0.005 \\
\hline
J1555$-$2908 & 2017-02-17 & 3606 & 238.92083 & -44.09250 &  &  \\
\hline
J1611$-$6011 & 2015-10-18 & 3606 & 242.80833 & -60.19833 & J1611.6$-$6013 & 0.059 \\
\hline
J1623$-$6935 & 2016-02-03 & 3606 & 245.95833 & -69.58583 & J1623.9$-$6936 & 0.023 \\
\hline
J1639$-$5146 & 2017-02-16 & 3606 & 249.85000 & -51.77611 & J1639.3$-$5146 & 0.007 \\
 & 2017-02-22 & 3606 &  &  &  &  \\
 & 2017-03-19 & 3606 &  &  &  &  \\
\hline
J1643$-$3148 & 2015-10-18 & 3606 & 250.85000 & -31.81472 & J1643.3$-$3148 & 0.008 \\
 & 2016-02-04 & 3606 &  &  &  &  \\
\hline
J1646$-$4405 & 2017-02-16 & 3606 & 251.59583 & -44.09250 & J1646.5$-$4406 & 0.024 \\
 & 2017-02-22 & 987 &  &  &  &  \\
 & 2017-03-19 & 3606 &  &  &  &  \\
\hline
J1649$-$3010 & 2016-11-05 & 3606 & 252.45417 & -30.18194 & J1649.8$-$3010 & 0.003 \\
\hline
J1653$-$0158 & 2016-11-05 & 3606 & 253.40000 & -1.97806 & J1653.6$-$0158 & 0.004 \\
\hline
J1656$-$2733 & 2016-11-05 & 2407 & 254.15000 & -27.55694 & J1656.5$-$2733 & 0.007 \\
\hline
J1704$-$4856 & 2016-02-03 & 3606 & 256.02500 & -48.94583 & J1705.4$-$4850 & 0.251 \\
 & 2016-03-28 & 3606 &  &  &  &  \\
\hline
J1711$-$3004 & 2015-10-17 & 2983 & 257.78750 & -30.06778 & J1711.0$-$3002 & 0.024 \\
\hline
J1715$-$3324 & 2016-11-05 & 3606 & 258.75000 & -33.41222 & J1714.9$-$3324 & 0.010 \\
 & 2017-02-22 & 3606 &  &  &  &  \\
 & 2017-03-19 & 3606 &  &  &  &  \\
\hline
J1717$-$5804 & 2017-02-16 & 3606 & 259.39167 & -58.07417 & J1717.5$-$5804 & 0.008 \\
\hline
J1721$-$5255 & 2016-11-06 & 3606 & 260.29583 & -52.92306 & J1721.3$-$5257 & 0.043 \\
\hline
J1739$-$2530 & 2017-02-17 & 3606 & 264.87917 & -25.50472 & J1739.3$-$2531 & 0.044 \\
\hline
J1743$-$4322 & 2016-11-06 & 3606 & 265.96667 & -43.38167 & J1743.7$-$4321 & 0.032 \\
\hline
J1747$-$3505 & 2017-02-16 & 3606 & 266.75000 & -35.09833 & J1747.0$-$3505 & 0.017 \\
\hline
J1759$-$3849 & 2016-11-06 & 3606 & 269.80833 & -38.82278 & J1759.1$-$3849 & 0.012 \\
 & 2017-02-13 & 3606 &  &  &  &  \\
\hline
J1802$-$4718 & 2016-02-03 & 3606 & 270.71250 & -47.30806 & J1802.8$-$4719 & 0.011 \\
 & 2016-03-28 & 1806 &  &  &  &  \\
 %& 2016-11-04 & 3606 &  &  &  &  \\
 %& 2016-11-06 & 3606 &  &  &  &  \\
 %& 2017-02-17 & 3606 &  &  &  &  \\
\hline
J1805$-$3619 & 2015-10-17 & 2407 & 271.25833 & -36.33056 & J1805.1$-$3618 & 0.029 \\
 & 2015-11-12 & 646 &  &  &  &  \\
\hline
J1812$-$3144 & 2016-02-03 & 3606 & 273.20833 & -31.73583 & J1812.8$-$3144 & 0.012 \\
 & 2016-03-28 & 3606 &  &  &  &  \\
\hline
J1813$-$6847 & 2017-02-16 & 3606 & 273.39167 & -68.78722 & J1813.7$-$6846 & 0.022 \\
\hline
J1816$-$6405 & 2016-11-06 & 3606 & 274.12083 & -64.09861 & J1816.4$-$6405 & 0.014 \\
 & 2017-02-13 & 643 &  &  &  &  \\
 & 2017-02-17 & 3606 &  &  &  &  \\
\hline
J1818$-$3333 & 2015-11-22 & 3606 & 274.50833 & -33.56444 & J1817.9$-$3334 & 0.025 \\
\hline
J1822$-$4717 & 2016-11-06 & 3606 & 275.74167 & -47.29750 & J1822.9$-$4718 & 0.008 \\
 & 2017-02-17 & 3606 &  &  &  &  \\
\hline
J1823$-$3543 & 2015-10-18 & 3606 & 275.96250 & -35.73278 & J1823.8$-$3544 & 0.009 \\
\hline
J1824$-$5426 & 2016-11-06 & 3606 & 276.06250 & -54.43333 & J1824.2$-$5427 & 0.019 \\
 & 2017-02-17 & 3606 &  &  &  &  \\
\hline
J1832$-$3840 & 2015-10-17 & 1888 & 278.23750 & -38.67556 & J1833.0$-$3840 & 0.019 \\
 %& 2015-11-12 & 3606 &  &  &  &  \\
 %& 2016-01-14 & 2057 &  &  &  &  \\
 %& 2016-01-17 & 3606 &  &  &  &  \\
 %& 2016-01-17 & 3606 &  &  &  &  \\
 %& 2016-03-28 & 1806 &  &  &  &  \\
 %& 2016-11-04 & 3606 &  &  &  &  \\
 %& 2016-11-06 & 2706 &  &  &  &  \\
 %& 2017-02-17 & 2706 &  &  &  &  \\
\hline
J1844$-$3840 & 2017-02-17 & 3606 & 281.15000 & -38.68056 &  &  \\
\hline
J1858$-$5425 & 2016-11-06 & 3606 & 284.62083 & -54.41778 & J1858.3$-$5424 & 0.027 \\
\hline
J1911$-$4857 & 2015-10-17 & 2856 & 287.86667 & -48.95806 & J1911.4$-$4856 & 0.012 \\
 & 2015-11-22 & 3606 &  &  &  &  \\
 & 2016-01-17 & 3606 &  &  &  &  \\
\hline
J1956$-$7012 & 2016-11-06 & 3606 & 299.12500 & -70.20833 & J1956.6$-$7011 & 0.025 \\
\hline
J2000$-$0258 & 2017-02-17 & 2603 & 300.22083 & -2.96750 &  &  \\
\hline
J2029$-$4237 & 2016-11-06 & 3606 & 307.37917 & -42.63000 & J2029.5$-$4237 & 0.002 \\
 & 2017-02-17 & 3606 &  &  &  &  \\
\hline
J2045$-$6835 & 2016-11-06 & 3606 & 311.44583 & -68.59917 & J2045.9$-$6835 & 0.016 \\
 & 2017-02-17 & 3606 &  &  &  &  \\
\hline
J2219$-$6837 & 2016-11-06 & 3606 & 334.95833 & -68.61889 & J2219.7$-$6837 & 0.004 \\
\hline
J2342$-$4747 & 2017-02-17 & 3606 & 355.69583 & -47.78889 & J2342.4$-$4739 & 0.137 \\
 & 2017-02-20 & 3606 &  &  &  &  \\
 & 2017-03-19 & 3606 &  &  &  &  \\
\hline
J2355$-$5247 & 2016-11-06 & 3606 & 358.79583 & -52.79389 & J2355.2$-$5247 & 0.006 \\
 & 2017-02-13 & 1303 &  &  &  &  \\
 & 2017-02-17 & 3606 &  &  &  &  \\
\hline
J2355$-$6612 & 2016-11-06 & 3606 & 358.91667 & -66.20167 & J2355.5$-$6614 & 0.034 \\
 & 2017-02-20 & 3606 &  &  &  &  \\
 & 2017-03-19 & 3606 &  &  &  &  \\
\enddata
%\tablecomments{4FGL-DR4 counterparts are determined from the closest catalog entry.  When no entry lies within 0.3\arcdeg, no counterpart is listed.  A transcription error in the position of J1555$-$2908 was made.}
\label{tab:targets}
\end{deluxetable*}